\documentclass[sn-mathphys,Numbered]{sn-jnl}
\usepackage{graphicx}%
\usepackage{multirow}%
\usepackage{amsmath,amssymb,amsfonts}%
\usepackage{amsthm}%
\usepackage{mathrsfs}%
\usepackage[title]{appendix}%
\usepackage{xcolor}%
\usepackage{textcomp}%
\usepackage{manyfoot}%
\usepackage{booktabs}%
\usepackage{algorithm}%
\usepackage{algorithmicx}%
\usepackage{algpseudocode}%

\begin{document}

\title{Asymptotically Optimal prepare-measure Quantum Key Distribution Protocol}

\author*[1,2]{\fnm{Hao} \sur{Shu}}\email{Hao\_B\_Shu@163.com}

\affil*[1]{\orgname{Shenzhen University}}

\affil[2]{\orgname{South China University of Technology}}

\abstract{Quantum key distribution (QKD) could be the most significant application of quantum information theory. In nearly four decades, although substantial QKD protocols are developed, the BB84 protocol and its variants are still the most researched ones. It is well-known that the secure bound of qubit error rate (QBER) of BB84 protocol is about 11$\%$ while it can be increased to 12.6$\%$ by six-state protocol. It would not be surprising that employing more basis could increase the bound. However, what is the optimal protocol, and how to analyze it? In this paper, investigations of asymptotically optimal QKD protocols are proposed. Precisely, We present an abstraction of prepare-measure QKD protocols and investigate two special cases which are optimal among all protocols coding by the same states. Our analysis demonstrates that the asymptotically optimal QBER bounds coding by orthogonal qubits are about 27.28$\%$ for both memory C-NOT attacks and memoryless C-NOT attacks while the bounds coding by non-orthogonal states in two mutually unbiased bases are about 22.73$\%$ for memory and 28.69$\%$ for memoryless C-NOT attacks. The protocols are idealized but might be asymptotically realized while their optimality indicates the ultimate potential of QKD protocols. Although the analysis only contains a special kind of attack, it provides a framework for investigating such protocols.}


\maketitle

\section{Introduction}

Communicating securely is always one of the most important fields in information theory. Nowadays, the only scheme whose security has been proven is coding with a one-time pad, which, however, could not be distributed to separated partners securely by classical methods. Therefore, quantum key distribution (QKD), a kind of scheme for distributing a one-time pad by quantum methods with security only depending on physical laws, becomes significant.

The first QKD protocol, called BB84 protocol, was proposed in 1984\cite{BB1984Quantum}, of which the security has been proven\cite{SP2000Simple}. In nearly four decades, substantial QKD protocols are developed\cite{E1991Quantum,CB2002Security,K2006A,LC1999Unconditional,S2021Quantum,ST2016A,GR2010Quantum,S2022Measurement} but BB84 protocol and its variants such as B92 protocol\cite{B1992Quantum}, BBM92 protocol\cite{BB1992Quantum}, six-state protocol\cite{B1998Optimal}, SARG04 protocol\cite{SA2004Quantum} and others\cite{SP2000Simple} are still the most researched ones.

Generally speaking, the security of QKD protocols comes from that if there is an eavesdropper who obtains enough information about the secret key, then she will create enough errors that are detectable by the legitimate partner. However, in practically implementing a QKD protocol, errors are unavoidable due to the imperfections of channels and devices. Therefore, to analyze the security of a QKD protocol, we have to estimate the threshold (or called secure bound) of the qubit error rate (QBER) it can tolerate, namely the value that the legitimate partner can extract a secret key by error-correcting and privacy amplification procedure when the QBER is below it\footnote{The simplest examples of error-correction as well as privacy amplification procedures might be employing XOR operations (although practically there might be substantial ways that are better). For the error-correcting procedure, assume that the legitimate partner, Alice and Bob, shares a one-time pad with probability $p<50\%$ for each bit in Bob's pad does not agree with Alice's. Then Alice can randomly choose two bits, implement the XOR operation (adding them in the mod2 sense), and publish the result together with the positions of the bits (but not what the bits are) publicly. Bob then chooses the bits in the same position, implementing XOR and comparing his results with Alice's. The two bits are discarded if the XOR results are different while one of the two bits is discarded with the other is remain employed if the XOR results are the same. Hence, the disagree rate of the left bit is $p^{2}$. By repeating the procedure, the legitimate could reduce the error rate to an acceptable level with a shorter key. Similarly for the privacy amplification procedure, assume that the probability for the eavesdropper, Eve, knows a bit of Alice and Bob is $q$. Then Alice can randomly choose two bits, implement the XOR operation and publicly publish the positions of the bits without the XOR result. Bob then chooses the bits in the same positions and implements XOR. The ordinary bits are discarded and the XOR result will be the new bit with the probability $q^{2}$ known by Eve.}.

There are works analyzing the security of protocols\cite{B2006Eavesdropping,B1998Optimal,P2008Symmetric,PA2019Advances,SR1998Security,N1996Security,FG1997Optimal,BA2011Optimal,BG1999Incoherent}, demonstrating that the threshold of QBER for BB84 protocol is about 11$\%$ while it is increased to about 12.6$\%$ for six-state protocol, under individual attacks. For memoryless attacks, the bound is about 15.4$\%$ for BB84 protocol, 20.4$\%$ for six-state protocol, and 17.6$\%$ for SARG04 protocol\cite{N1996Security,BA2011Optimal}.

It would not be surprising that a protocol employing more basis for coding should be more secure. However, what is the optimal one, namely what protocol is the most secure one, theoretically, and how to analyze it? In this paper, We present an abstraction of prepare-measure QKD protocols and investigate two special cases which are optimal among all protocols coding by the same states. We calculate the secure QBER bounds for the special protocols under C-NOT attacks, demonstrating that the asymptotically optimal QBER bounds coding by orthogonal qubits are about 27.28$\%$ for both memory and memoryless C-NOT attacks while the bounds coding by non-orthogonal qubits in two mutually unbiased bases are increased to about 22.73$\%$ for memory and 28.69$\%$ for memoryless C-NOT attacks. Our investigations also reveal the meaningless of collective C-NOT attacks, namely employing C-NOT attack to every qubit. The optimality of our protocols indicates the ultimate potential of security under such attacks. Although the above attacks might not represent the general one, our analysis provides a framework for analyzing the abstraction protocol. Finally, despite their idealization, our protocols might be realized asymptotically.

\section{The abstraction of general prepare-measure QKD protocol}

For simplicity, our scenario is under two assumptions.

(1) The legitimate partner, Alice and Bob, can employ quantum memories.

(2) Alice and Bob employ noiseless channels including side channels such as measurements and single-photon sources in which the eavesdropper, Eve, will not employ attacks based on photon numbers.

The abstraction of a general prepare-measure protocol is described as follows.
\\

\textbf{Protocol:}
\\

\textbf{Step 1}: Alice and Bob agree to encode 0 by state $C_{0}|0\rangle$ and 1 by state $C_{1}|0\rangle$, where $C_{i}, i=1, 2$ are unitary operators on qubits. The states can be orthogonal or non-orthogonal. Alice chooses a bit string randomly and for each bit, she chooses a unitary operator, $U$ (depending on the special protocol and can be randomly in a set), and sends $UC_{i}|0\rangle$ with $i=1, 2$ chosen randomly, to Bob.

\textbf{Step 2}: After Bob receives the state, Alice publicly announces the choice of $U$. Bob measures the qubit via basis $U_{B}|0\rangle, U_{B}|1\rangle$, where $U_{B}$ depends on the chosen protocol (and $U$), to decode the bit.

These steps will be repeated several times until Alice and Bob share a long enough bit string.

\textbf{Step 3}: Alice and Bob discard the non-effective bits (depending on the chosen protocol) and estimate the QBER by declaring part of their bit string and public discussions. The string is aborted if the QBER is too high or not random enough.

\textbf{Step 4}: If the error rate and the randomness of the string are acceptable, they generate a raw secret key by remaining bits.
\\

Certainly, standard post-processing procedures such as error-correcting and privacy amplification are needed to transfer the raw key to an employable one, but they are done classically\footnote{One can view that the QKD protocol is already finished once a raw key is obtained since the post-processing procedures are classical, but should note that such procedures are necessary for a final key. The classical parts are usually viewed as an independent research direction and have a history longer than QKD. For now, one can just note that they are available under conditions.}.

Note that if $C_{0}$ is chosen to be $I$, $C_{1}$ is chosen to be the Pauli operator $X$, $U$ is chosen as $I$ or Hadamard gate $H$ randomly for each bit, and $U_{B}$ is chosen as $U$, then the protocol becomes BB84 protocol with Hadamard gates\cite{SP2000Simple}, while if $C_{0}=I$, $C_{1}=H$, $U=I$ and $U_{B}$ is chosen as $I$ or $H$ randomly for each bit, then the protocol becomes B92 protocol\cite{B1992Quantum}. Also if $C_{0}$ is chosen randomly among $I$ and $X$ for each bit, $C_{1}$ is chosen randomly among $H$ and $HX$ for each bit, $U=I$ and $U_{B}$ is chosen as $I$ or $H$ randomly for each bit, then the protocol becomes SARG04 protocol\cite{SA2004Quantum} by Alice declares whether $C_{i}\in\{I, H\}$ for each $C_{i}|0\rangle$ she sent after Bob's receiving while if $C_{0}$ is randomly chosen in $\{I, H\}$ for each bit, $C_{1}$ is randomly chosen in $\{X, HX\}$ for each bit, $U=I$ and $U_{B}$ is randomly chosen in $\{I, H\}$ for each bit, then the protocol becomes ordinary BB84 protocol\cite{BB1984Quantum}.

\section{Two special protocols and the optimality}

We would like to investigate two special protocols. The first one chooses $C_{0}=I, C_{1}=X, U_{B}=U$ while $U$ is chosen randomly among all unitary operators on qubits for each bit. Therefore, the protocol expands the BB84 protocol. We will call it BB84 type protocol. The second one chooses $C_{0}=I, C_{1}=H$, $U_{B}$ be $UI$ or $UH$ randomly for each bit while $U$ is chosen randomly among all unitary operators on qubits for each bit. Therefore, the protocol expands the B92 protocol. We will call it the B92 type protocol. Similarly, we can have SARG04 type protocol\footnote{The choice should be: $C_{0}$ randomly among $I$ and $X$, $C_{1}$ randomly among $H$ and $HX$, $U$ randomly among all unitary operators on qubits and $U_{B}$ randomly in $UI$ or $UH$. And Alice needs to declare whether $C_{i}\in\{I, H\}$ for each $UC_{i}|0\rangle$ she sent after Bob's receiving. However, there is no difference between Alice chooses $C_{i}$ in $\{I, H\}$ or in $\{X, HX\}$, since the randomness of $U$ implies the randomness of $UXZ$, where $Z$ is the Pauli operator $Z$ and $XZ$ transforms $\{|0\rangle, |+\rangle\}$ to $\{|1\rangle, |-\rangle\}$ upon a global phases. Moreover, in such a perspective, Alice does not need to declare whether $C_{i}\in\{I, H\}$, since it always does, and it would not benefit Eve since the state is random for her before received by Bob while whether $C_{i}\in\{I, H\}$ makes no difference after Bob receives the state.}. However, the investigations are exactly the same as the ones in the B92 type and thus we only investigate the B92 type\footnote{The difference between the B92 protocol and the SARG04 protocol is detailed as follows. In the B92 protocol, Alice sends a state in $\{|0\rangle,|+\rangle\}$ randomly for each bit while Bob measures via one of the BB84 basis. If Bob obtains an outcome in $\{|1\rangle,|-\rangle\}$, then he knows the bit sent by Alice and the bit is effective, otherwise they discard the bit. In contrast, in SARG04 protocol, Alice sends a state in $S_{1}=\{|0\rangle,|+\rangle\}$ or $S_{2}=\{|1\rangle,|-\rangle\}$ randomly for each bit while Bob measures via one of the BB84 basis. After Bob receives the state, Alice declares whether the state belongs $S_{1}$ or $S_{2}$, Bob can determine the bit of Alice if either Alice chose $S_{1}$ and he obtains $|1\rangle$ or $|-\rangle$, or Alice chose $S_{2}$ and he obtains $|0\rangle$ or $|+\rangle$, while the bit is discard otherwise. 
	
	Hence, if a random $U$ is implemented on the sent state, then B92-type and SARG04-type protocols have no difference before Bob receives since the sent state is random for anyone but Alice, and after Bob receives the state and Alice declares her choice, B92 and SARG04 protocols are the same. Therefore, the two types are equivalent.}.

Our protocols are optimal among all prepare-mesure protocols coding with the same states (namely coding 0 by state $C_{0}|0\rangle$ and 1 by state $C_{1}|0\rangle$). The optimality can be demonstrated as follows. Whatever Alice sends, since Eve is assumed to have any technology under physical laws, she can (1) randomly operate the state by a unitary operator $U$, followed by (2) implement a normal attack, and finally (3) operate $U^{\dagger}$ on the partita (in this paper, a partita always represents a subsystem) sent to Bob. Such operations of Eve allow the whole procedure equal to that Alice and Bob implement our protocol while Eve implements the normal attack. In other words, Eve would be able to transform any protocol into ours if it provides fewer benefits for her and then normally implement attacks. Therefore, for Eve, our protocol could provide the least benefit. Hence, our protocol is the worst one for Eve and thus the optimal one for Alice and Bob.

\section{Attack of Eve}

The aim of Eve is to guess the bits of Alice correctly as many as possible without resulting in the abortion of the protocol. Here we would discuss a kind of individual attack that Eve copies a qubit sent by Alice with a C-NOT gate, called a C-NOT attack and also known as a probe-entangling attack discussed in some previous protocols\cite{SR1998Security,B2003Optimum,B2005Quantum,SW2006Attacking,S2006Performance}.

The C-NOT attacks can be described as follows. For a state sent by Alice, Eve adds an auxiliary partita (her auxiliary system) and operates a C-NOT gate under a chosen basis. Then she sends the ordinary partita (qubit) to Bob while storages the auxiliary one. In the memory case, Eve's state is measured individually after she eavesdropped on all classical communications of Alice and Bob, and her measurement depends on classical messages she obtained, while in the memoryless case, Eve measures her state immediately.

Assume that the state sent by Alice is $U|c\rangle=aE'|0\rangle_{A}+bE'|1\rangle_{A}$. After Eve's action, the state becomes $|X_{c}\rangle=aE'|0\rangle_{A}|0\rangle_{E}+bE'|1\rangle_{A}|1\rangle_{E}$\footnote{In details, assume that Alice encodes states $|c\rangle$ by operator $U$ and thus sends state $U|c\rangle$ to Bob while the C-NOT attack provided by Eve is related to basis $\{|0_{E}\rangle, |1_{E}\rangle\}$. By orthonormality, there is a unitary operator $E'$ such that $|0_{E}\rangle=E'|0\rangle$, and $|1_{E}\rangle=E'|1\rangle$. Rewrite $U|c\rangle$ in the basis $\{|0_{E}\rangle, |1_{E}\rangle\}$ as $U|c\rangle=a|0_{E}\rangle+b|1_{E}\rangle$, where the coefficients are calculated as $a=\langle 0_{E}|U|c\rangle$,$b=\langle 1_{E}|U|c\rangle$. The attack of Eve is as follows. (1) Eve employs an auxiliary partita, making the state be $U|c\rangle|0_{E}\rangle=a|0_{E}\rangle_{A}|0_{E}\rangle_{E}+b|1_{E}\rangle_{A}|0_{E}\rangle_{E}$, (2) Eve implements a C-NOT gate, making the state be $|X_{c}\rangle=a|0_{E}\rangle_{A}|0_{E}\rangle_{E}+b|1_{E}\rangle_{A}|1_{E}\rangle_{E}=aE'|0\rangle_{A}|0_{E}\rangle_{E}+bE'|1\rangle_{A}|1_{E}\rangle_{E}$. Finally, since $|0_{E}\rangle$ and $|1_{E}\rangle$ only represents two orthonormal states of Eve, we rewrite it as $\{|0\rangle, |1\rangle\}$ for simplity, and the state becomes $|X_{c}\rangle=aE'|0\rangle_{A}|0\rangle_{E}+bE'|1\rangle_{A}|1\rangle_{E}$.}, where $E$ denotes the partita of Eve, $E'$ is a unitary operator on qubits, $|0\rangle_{E}$, $|1\rangle_{E}$ are two orthogonal states in $E$ (the dimension of $E$ could be larger) and $a=\langle 0|E'^{\dagger}U|c\rangle$,$b=\langle 1|E'^{\dagger}U|c\rangle$.

\section{Secure bound of QBER}

The secure condition for the legitimate partner is the allowance to extract a secret key, which is promised by the private information being larger than zero. The private information of the legitimate partner also provides the secret key rate\footnote{The private information depends on the QBER of both the legitimate partner and the eavesdropper. Some schemes in key distillation can be found in \cite{GL2004Security,BB1988Privacy,BB1992Experimental}.}\cite{M1993Secret,GL2004Security,KG2005Lower,SB2009The}.

Denote the QBER that Alice and Bob decide to tolerate by $r$ and the error rate of them when Eve attacks a state by $e_{B}$. If Alice and Bob obtain $N$ bits in which $t$ bits are attacked, then $e_{B}t\leq rN$ for not resulting in the abortion of the string. Therefore, the proportion of qubits Eve can attack is at most $\frac{r}{e_{B}}$. An easy discussion shows that in an optimal strategy of Eve, $\frac{r}{e_{B}}\leq 1$. Now, the private information is calculated (under assumptions that $\frac{r}{e_{B}}\leq 1$ and Eve attacks $\frac{r}{e_{B}}$ states such that the average error rate in all bits of Bob is $r$) as follows, where $e$ is the error rate of Eve (namely the probability of Eve of guessing a bit wrongly) when she launches an attack.

\begin{equation}
	\begin{aligned}	
		I(A:B)-I(A:E)&=I(A:B)-\frac{r}{e_{B}}I(A:E)_{att}
		\\
		&=1-h(r)-\frac{r}{e_{B}}(1-h(e)).	
	\end{aligned}
\end{equation}

\noindent where $h(x)=-xlogx-(1-x)log(1-x)$ is the binary entropy and the subscript $'att'$ represents the case when Eve attacks. Here, we assume that the error rate of Eve on bits 0 and bits 1 are the same, which is not surprising since Alice and Bob code with symmetric states (in our discussions below, Alice and Bob always employ symmetric states for coding) and thus if an optimal strategy of Eve obtains more errors on bits 0, then she can employ a symmetric strategy, obtaining more errors on bits 1 and she can combine the strategies (still be optimal) such that the error rate on bits 0 and bits 1 are the same\footnote{In fact, since $h(x)$ is convex, the equality of error rate would benefit Eve mostly, providing the fixed average error rate.}.

\section{Asymptotically optimal QBER bound of BB84 type protocol}

Let us investigate the BB84 type protocol. In such a protocol, Alice sends state $|0\rangle$ or $|1\rangle$ operated by $U$ randomly while Bob measures via basis $U|0\rangle$, $U|1\rangle$. We shall calculate $e$ and $e_{B}$. As shown in section IV, after Eve's attack, the state sent by Alice becomes $|X_{c}\rangle=aE'|0\rangle_{A}|0\rangle_{E}+bE'|1\rangle_{A}|1\rangle_{E}$, where $c=0, 1$. Write $U=\begin{pmatrix}
	u_{1}	& u_{2}\\
	u_{3}	& u_{4}
\end{pmatrix}$ under the computational basis,
and since $U$ is unitary, $|u_{1}|^{2}+|u_{2}|^{2}=1, |u_{1}|=|u_{4}|, |u_{2}|=|u_{3}|, u_{1}\bar{u_{3}}+u_{2}\bar{u_{4}}=0$. Assume that Eve measures her partita by the positive operator-valued measurement (POVM) $\{M, N\}$ (note that for general attacks, an optimal attack for Eve can contain only two measurement outcomes since she only guesses the bit of Alice and Bob be 0 or 1), depending (for memory attacks) or not depending (for memoryless attacks) on $U$.

\subsection{memory C-NOT attack}

In the situation that Eve launches a memory C-NOT attack, both Bob and Eve measure their states after knowing $U$. Now,

\begin{equation}
	\begin{aligned}
		P_{B|A}(0|0)&
		=\int_{U,ave}\langle X_{0}|(U|0\rangle \langle 0|U^{\dagger}\otimes I)|X_{0}\rangle dU
		\\
		&=\int_{U,ave}(|\langle 0|E'^{\dagger}U|0\rangle|^{2}|\langle 0|E'^{\dagger}U|0\rangle|^{2}+|\langle 1|E'^{\dagger}U|0\rangle|^{2}|\langle 1|E'^{\dagger}U|0\rangle|^{2}) dU
		\\
		&=\int_{U,ave}|\langle 0|U|0\rangle|^{2}|\langle 0|U|0\rangle|^{2} dU+\int_{U,ave}|\langle 1|U|0\rangle|^{2}|\langle 1|U|0\rangle|^{2} dU
		\\
		&=2\int_{U,ave}|\langle 0|U|0\rangle|^{4} dU
		=2\int_{U,ave}|u_{1}|^{4} dU
		\\
		&=2\int_{S:|u_{1}|^{2}\leq 1, u_{1}\in C,ave}|u_{1}|^{4} du_{1}=2\int_{S:x^{2}+y^{2}\leq 1, x,y\in R,ave}(x^{x}+y^{2})^{2} dxdy=\frac{2}{3},
	\end{aligned}
\end{equation}

where $\int_{U,ave}$\footnote{Strictly speaking, the measure of unitary operators should be clarified. However, for simplicity, this would not be done in the paper but note that all integrals in the paper can be viewed as calculating in the complex plane and further viewed as in the real plane. For instance, randomly sampling unitary operator $U$ in $\int |u_{1}|^{4}$ can be done by randomly sampling $u_{1}$ inside the unit cycle of the complex plane, and then be viewed as inside the unit cycle of the real plane.} represents integrating over all unitary operators and taking the average (namely, divided by $\int_{U}1 dU$). The calculation gives $P_{B|A}(1|1)=\frac{2}{3}$ and $e_{B}=\frac{1}{3}$.

To calculate $e$, without loss generality, assume that Eve will guess the bit of Alice and Bob to be 0 if her measurement outcome is $M$ and 1 if her outcome is $N$.

\begin{equation}
	\begin{aligned}
		&P_{E|A}(0|0)=P_{B|A}(0|0)P_{E|A,B}(0|0,0)+P_{B|A}(1|0)P_{E|A,B}(0|0,1)
		\\
		=&\int_{U,ave}\langle X_{0}|(U|0\rangle \langle 0|U^{\dagger}\otimes I)(I\otimes M_{U})(U|0\rangle \langle 0|U^{\dagger}\otimes I)|X_{0}\rangle dU
		\\
		&+\int_{U,ave}\langle X_{0}|(U|1\rangle \langle 1|U^{\dagger}\otimes I)(I\otimes M_{U})(U|1\rangle \langle 1|U^{\dagger}\otimes I)|X_{0}\rangle dU
		\\
		=&\int_{U,ave}\langle X_{0}|(I\otimes M_{U})|X_{0}\rangle dU
		\\
		=&\int_{U,ave}(|\langle 0|E'^{\dagger}U|0\rangle|^{2} \langle 0|M_{U}|0\rangle+|\langle 1|E'^{\dagger}U|0\rangle|^{2} \langle 1|M_{U}|1\rangle) dU
		\\
		=&\int_{U,ave}(|\langle 0|U|0\rangle|^{2} \langle 0|M_{E'U}|0\rangle+|\langle 1|U|0\rangle|^{2} \langle 1|M_{E'U}|1\rangle) dU.
	\end{aligned}
\end{equation}

Similarly, $P_{E|A}(1|1)=\int_{U,ave}(|\langle 0|U|1\rangle|^{2} \langle 0|N_{E'U}|0\rangle+|\langle 1|U|1\rangle|^{2} \langle 1|N_{E'U}|1\rangle) dU$. Now

\begin{equation}
	\begin{aligned}
		P_{E|A}&(0|0)+P_{E|A}(1|1)
		\\
		=&\int_{U,ave}(|\langle 0|U|0\rangle|^{2} \langle 0|M_{E'U}|0\rangle
		+|\langle 1|U|0\rangle|^{2}
		\langle 1|M_{E'U}|1\rangle) dU
		\\
		&+\int_{U,ave}(|\langle 0|U|1\rangle|^{2} \langle 0|N_{E'U}|0\rangle+|\langle 1|U|1\rangle|^{2} \langle 1|N_{E'U}|1\rangle) dU
		\\
		=&\int_{U,ave}(|u_{1}|^{2} \langle 0|M_{E'U}|0\rangle+|u_{2}|^{2} \langle 1|M_{E'U}|1\rangle
		+|u_{2}|^{2} \langle 0|N_{E'U}|0\rangle+|u_{1}|^{2} \langle 1|N_{E'U}|1\rangle) dU
		\\
		\leq&2\int_{U,ave}max(|u_{1}|^{2},|u_{2}|^{2}) dU=2\int_{U,|u_{1}|\geq |u_{2}|, |u_{1}|^{2}+|u_{2}|^{2}=1, ave}|u_{1}|^{2} dU
		\\
		&=2\int_{S:\frac{1}{2}\leq |u_{1}|^{2}\leq 1, u_{1}\in C, ave}|u_{1}|^{2} du_{1}=2\int_{S:\frac{1}{2}\leq x^{2}+y^{2}\leq 1, x,y\in R, ave}(x^{2}+y^{2}) dxdy=\frac{3}{2}.
	\end{aligned}
\end{equation}

Note that since the error rate of Bob(namely $e_{B}$) is definitized, an optimal strategy of Eve is making $P_{E|A}(0|0)+P_{E|A}(1|1)$ maximal, which represents that she guesses correctly in the most bits, and can be done by choosing her partita to be $C^{2}$ with $M_{EU}=\begin{pmatrix}
	1	& 0\\
	0	& 0
\end{pmatrix},
N_{EU}=\begin{pmatrix}
	0	& 0\\
	0	& 1
\end{pmatrix}$
if $|u_{1}|\geq|u_{2}|$ and
$M_{EU}=\begin{pmatrix}
	0	& 0\\
	0	& 1
\end{pmatrix},
N_{EU}=\begin{pmatrix}
	1	& 0\\
	0	& 0
\end{pmatrix}$
otherwise, under the computational basis. It is not surprising that an optimal strategy of Eve is employing $C^{2}$ as her auxiliary partita since extending the dimension of the system would not give benefits on distinguishing states\cite{S2020The}. Hence, under the optimal attack of Eve, $P_{E|A}(0|0)=\frac{1}{2}(P_{E|A}(0|0)+P_{E|A}(1|1))=\frac{3}{4}$ and thus $e=\frac{1}{4}$. Finally, the secure threshold of QBER is about $\approx 27.28\%$.

It is worth noting that if Eve's measurement is restricted to projective ones, then $P_{E|A}(0|0)$ and $P_{E|A}(1|1)$ can be calculated directly and become maximal in the same strategy, which is coincident with the result above. Hence, an optimal strategy for Eve can only employ projective measurements for measuring.

\subsection{Memoryless attack}

In the situation that Eve launches a memoryless attack, Eve measures the state firstly without knowing $U$. Now,

\begin{equation}
	\begin{aligned}
		P_{B|A}(0|0)
		=&P_{E|A}(M|0)P_{B|A,E}(0|0,M)+P_{E|A}(N|0)P_{B|A,E}(0|0,N)
		\\
		=&\int_{U,ave}\langle X_{0}|(U|0\rangle \langle 0|U^{\dagger}\otimes M)|X_{0}\rangle dU+\int_{U,ave}\langle X_{0}|(U|0\rangle \langle 0|U^{\dagger}\otimes N)|X_{0}\rangle dU
		\\
		=&\int_{U,ave}\langle X_{0}|(U|0\rangle \langle 0|U^{\dagger}\otimes I)|X_{0}\rangle dU
		=\int_{U,ave}(|u_{1}|^{4}+|u_{3}|^{4}) dU
		=\frac{2}{3}.
	\end{aligned}
\end{equation}

\begin{equation}
	\begin{aligned}
		P_{E|A}(0|0)=&\frac{1}{2}(\int_{U,ave,\langle X_{0}|I\otimes M|X_{0}\rangle \geq \langle X_{0}|I\otimes N|X_{0}\rangle}\langle X_{0}|I\otimes M|X_{0}\rangle dU
		\\
		&+\int_{U,ave,\langle X_{0}|I\otimes M|X_{0}\rangle < \langle X_{0}|I\otimes N|X_{0}\rangle}\langle X_{0}|I\otimes N|X_{0}\rangle dU)
		\\
		=&\int_{U,ave}max(\langle X_{0}|I\otimes M|X_{0}\rangle, \langle X_{0}|I\otimes N|X_{0}\rangle dU),
		\\
		P_{E|A}(1|1)=&\frac{1}{2}(\int_{U,ave,\langle X_{0}|I\otimes M|X_{0}\rangle \geq \langle X_{0}|I\otimes N|X_{0}\rangle}\langle X_{1}|I\otimes N|X_{1}\rangle dU
		\\
		&+\int_{U,ave,\langle X_{0}|I\otimes M|X_{0}\rangle < \langle X_{0}|I\otimes N|X_{0}\rangle}\langle X_{1}|I\otimes M|X_{1}\rangle dU)
		\\
		=&\int_{U,ave}max(\langle X_{1}|I\otimes M|X_{1}\rangle, \langle X_{1}|I\otimes N|X_{1}\rangle dU).
	\end{aligned}
\end{equation}

\begin{equation}
	\begin{aligned}
		P_{E|A}&(0|0)+P_{E|A}(1|1)
		\\
		=&\int_{U,ave}max(\langle X_{0}|I\otimes M|X_{0}\rangle, \langle X_{0}|I\otimes N|X_{0}\rangle) dU+\int_{U,ave}max(\langle X_{1}|I\otimes M|X_{1}\rangle, \langle X_{1}|I\otimes N|X_{1}\rangle) dU
		\\
		=&\int_{U,ave}max(|u_{1}|^{2}\langle 0|M|0\rangle+|u_{3}|^{2}\langle 1|M|1\rangle+|u_{3}|^{2}\langle 0|N|0\rangle+|u_{1}|^{2}\langle 1|N|1\rangle,
		\\
		&\qquad \qquad \quad |u_{3}|^{2}\langle 0|M|0\rangle+|u_{1}|^{2}\langle 1|M|1\rangle+|u_{1}|^{2}\langle 0|N|0\rangle+|u_{3}|^{2}\langle 1|N|1\rangle)dU
		\\
		\leq& 2\int_{U,ave}max(|u_{1}|^{2},|u_{3}|^{2})dU=\frac{3}{2},
	\end{aligned}
\end{equation}

where $\langle X_{0}|I\otimes M|X_{0}\rangle \geq \langle X_{0}|I\otimes N|X_{0}\rangle$ represents the bit being more likely to be 0 when Eve obtains outcome $M$ and 1 when she obtains outcome $N$. An optimal strategy for Eve is making $P_{E|A}(0|0)+P_{E|A}(1|1)$ maximal (similar to above) and can be done by choosing her partita to be $C^{2}$ with
$M=\begin{pmatrix}
	1	& 0\\
	0	& 0
\end{pmatrix},
N=\begin{pmatrix}
	0	& 0\\
	0	& 1
\end{pmatrix}$.
Hence, $e=\frac{1}{4}$, $e_{B}=\frac{1}{3}$ and thus the secure QBER threshold is also about 27.28$\%$.

It is worth noting that the investigations above also show that in such a protocol, the error rate created by Eve is always $\frac{1}{3}$ if she attacks. Hence, if Eve employs collective attacks, then the error rate of Bob could never satisfy the secure threshold, which results in the abortion of the protocol. Therefore, discussing collective attacks would be meaningless.

\section{Asymptotic QBER bound of B92 type protocol}

Let us investigate the B92-type protocol. In such a protocol, Alice sends state $|0\rangle$ or $|+\rangle$ while Bob measures via basis $\{U|0\rangle, U|1\rangle\}$ or $\{U|+\rangle, U|-\rangle\}$ randomly. Others are the same as in BB84 type protocol and we will employ the same symbols. Note that our investigations also hold for SARG04 type protocol.
\\

\subsection{memory C-NOT attack}

Now,

\begin{equation}
	\begin{aligned}
		P_{B|A}(|-\rangle||0\rangle)=&\int_{U,ave}\langle X_{0}|(U|-\rangle \langle -|U^{\dagger}\otimes I)|X_{0}\rangle dU
		\\
		=&\frac{1}{2}\int_{U,ave} [ |u_{1}|^{2}(1-2Re(u_{1}\bar {u_{2}}))+|u_{3}|^{2}(1-2Re(u_{3}\bar{u_{4}}))] dU
		\\
		=& \frac{1}{2}[1-2\int_{U,ave} (|u_{1}|^{2}-|u_{3}|^{2})Re(u_{1}\bar {u_{2}})) dU]
		=\frac{1}{2},
		\\
		P_{B|A}(|1\rangle||0\rangle)=&\int_{U,ave}\langle X_{0}|(U|1\rangle \langle 1|U^{\dagger}\otimes I)|X_{0}\rangle dU=2\int_{U,ave} |u_{1}u_{2}|^{2} dU
		\\
		&=2\int_{S:|u_{1}|^{2}+|u_{2}|^{2}=1, u_{1}, u_{2}\in C,ave} |u_{1}u_{2}|^{2} du_{1}du_{2}
		\\
		&=2\int_{S:x^{2}+y^{2}\leq 1, x, y\in R,ave} (x^{2}+y^{2})(1-x^{2}-y^{2}) dxdy=\frac{1}{3}.	
	\end{aligned}
\end{equation}

Therefore, $P_{B|A}(0|0)=\frac{P_{B|A}(|-\rangle||0\rangle)}{P_{B|A}(|-\rangle||0\rangle)+P_{B|A}(|1\rangle||0\rangle)}=\frac{3}{5}$, and $e_{B}=\frac{2}{5}$

Instead of calculating $e$, we use the fact that $e\geq\frac{1}{2}-\frac{1}{2\sqrt{2}}$. Since we can view Eve's action as a cloning procedure in which she can not do better than a perfect clone. However, for a perfect clone, Eve has to distinguish two non-orthogonal states $|0\rangle$ and $|+\rangle$, which can be optimally distinguished with error rate $\frac{1}{2}-\frac{1}{2\sqrt{2}}$, and the proof is given in the supplied material. Hence, the secure bound of QBER is about $22.73\%$. Note that even if $e=0$, the secure bound would be about 15.30$\%$.

\subsection{Memoryless attack}

The investigation of memoryless attacks is similar to the above. $e_{B}$ is the same as in the memory attacks since there is no difference between Bob measures first and Eve measures first. $e$ is the same as in memoryless attacks of BB84 type. Hence, $e_{B}=\frac{2}{5}$ while $e=\frac{1}{4}$. The secure QBER threshold now becomes about $28.69\%$.

Similar to the BB84 type, note that the error rate created in a collective attack is always $\frac{2}{5}$, which could not satisfy the secure threshold. Therefore, collective attacks could always result in the abortion of the bit string. Hence, discussing collective attacks would also be meaningless.

\section{Discussion}

\subsection{Mutual information via distance}

\begin{figure}[ht]
	\centering
	\includegraphics[width=5in]{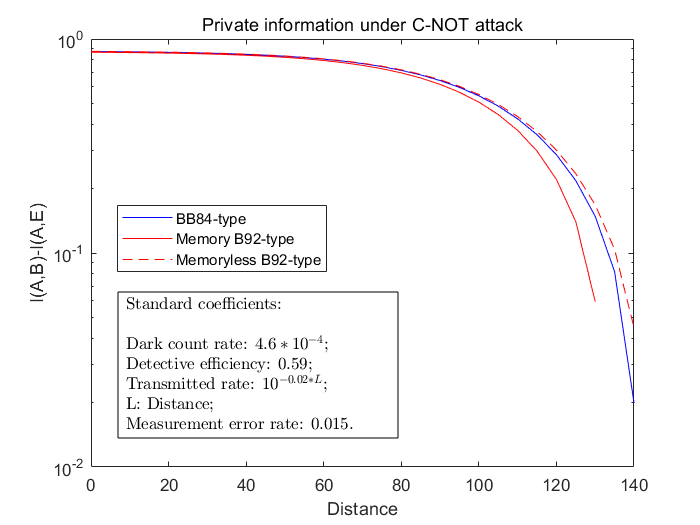}
	\caption{Private information via distance}
	\label{fig}
\end{figure}

We plot how the private information of the legitimate partner of the presented protocols decreases as transmitting distance increases under some settings, over C-NOT attacks. Please see Figure\ref{fig}.

\subsection{Asymptotically optimal}

Although our protocols are idealized and can not be practically implemented even with memories due to the infinite choices of $U$, the bounds can be asymptotically touched. To see this, just employ a finite number of $U$ which are uniformly distributed among all unitary operators on qubits. As the number of $U$ increases, the error rates and thus the QBER bounds can close to the above ones.

\subsection{Memoryless}

Also, the need for memories might be removed by sifting $U$ in a finite set similar to the basis sifting procedure in the BB84 protocol with Hadamard gates\cite{SP2000Simple}. However, if $n$ $U$ are employed without memories, the efficiency will be reduced to $\frac{1}{n}$ and approximates to 0 for large n. Therefore in practice, sifting with a large number of $U$ instead of employing a memory might be impractical, but with a small number of $U$ to improve the security might still be valid.

\subsection{Practical channels}

In practice, Alice and Bob might employ weak coherence sources and thus, Eve might employ, for example, photon-number-splitting (PNS) attacks\cite{HI1995Quantum,LJ2002Quantum}. However, these attacks are handled by other methods such as decoy state methods\cite{H2003Quantum,LM2005Decoy,MQ2005Practical}.

Also, practical channels might be lossy and measurements could be with errors. As a consequence, the error rate of Bob could climb as the distance increases (in lossy cases). The key rate could still depend on the private information but the error rates should be calculated together with these issues. However, these would not be discussed further in this paper.

\subsection{Other protocols}

The abstraction scheme could be extended to other protocols. In fact, all protocols involving the legitimate partner only such as BBM92 protocol\cite{BB1992Quantum}, which is based on entanglement, might apply random $U$ to improve the security.

\section{Conclusion}

In conclusion, we presented an abstraction of prepare-measure QKD protocols and investigated two special cases which are optimal among all protocols coding via the same states. The optimality indicates the ultimate potential of such protocols. For coding with orthogonal qubits (expanding BB84 protocol), we demonstrated that the optimal secure QBER bounds are about 27.28$\%$ for both memory and memoryless C-NOT attacks, while for coding with non-orthogonal qubits in two mutually unbiased bases (expanding B92 or SARG04 protocol), the secure bounds are increased to about 22.73$\%$ and 28.69$\%$ for memory and memoryless C-NOT attacks, respectively. We also demonstrated that an optimal strategy of Eve can only employ $C^{2}$ as her auxiliary partita and projective measurements for measuring and collective attacks are meaningless (in C-NOT attacks). Despite the idealization, our protocols could be asymptotically realized in memory cases and might provide improvements in memoryless cases. Finally, although the analysis in this paper only contains a special kind of attack, which is the most normal one but might not be the most general one, it provides a framework for investigating such protocols.

\section{Availability of data}

The data that supports the findings of this study are available within the article.

\section{Conflict of interest}

The author declares no conflict of interest.

\bibliography{Bibliog}

\section{Supplied material}

The optimal error rate of distinguishing states $|0\rangle$ and $|+\rangle$:

Suppose that one employs the positive operator-valued measurement (POVM) $\{M_{0}, M_{+}\}$ to distinguish the two states and judges the state to be $|0\rangle$ by outcome $M_{0}$ and $|+\rangle$ by outcome $M_{+}$. Write the operators as matrices under the computational basis as
$M_{0}=\begin{pmatrix}
	m_{1}       & m_{2}  \\
	\bar{m_{2}} & m_{4}
\end{pmatrix}$
$M_{+}=\begin{pmatrix}
	1-m_{1}       & -m_{2}  \\
	-\bar{m_{2}}  & 1-m_{4}
\end{pmatrix}$,
where $0\leq m_{1},m_{4}\leq 1$ are real and $m_{1}m_{4}\geq|m_{2}|^{2}, (1-m_{1})(1-m_{4})\geq|m_{2}|^{2}$, since $\{M_{0}, M_{+}\}$ is a POVM. Without loss generality, assume that $m_{1}+m_{4}\leq 1$. The correct rate is calculated as
\begin{equation}
	\begin{aligned}
		P_{correct}=\frac{1}{2}(\langle 0|M_{0}|0\rangle+\langle +|M_{+}|+\rangle)
		=\frac{1}{2}+\frac{1}{4}(m_{1}-m_{4}+2Re(m_{2}))
	\end{aligned}
\end{equation}
with conditions $0\leq m_{1},m_{4}\leq 1$ are real and $m_{1}m_{4}\geq|m_{2}|^{2}$. It is easy to see that to make $P_{correct}$ maximal, we can choose $m_{2}$ be real while if $m_{1}+m_{4} < 1$, we can enlarge $m_{1}$. Therefore, when calculating maximal $P_{correct}$, we can assume that $m_{1}+m_{4} = 1$ and $m_{2}$ is real. To maximise $P_{correct}$, we should assume that $m_{1}\geq m_{4}$. Hence, the problem becomes maximising $P_{correct}=\frac{1}{2}+\frac{1}{4}(2m_{1}+2m_{2}-1)$ in the area $\frac{1}{2}\leq m_{1}\leq 1, m_{1}-m_{1}^{2}\geq m_{2}^{2}$. For every $m_{1}$, to maximise $P_{correct}$, we should let $m_{2}$ as large as possible. Therefore, $m_{2}\geq 0$ and $m_{1}-m_{1}^{2}=m_{2}^{2}$. Then $P_{correct}=\frac{1}{2}+\frac{1}{4}(2m_{1}+2\sqrt{m_{1}-m_{1}^{2}}-1)$. When $m_{1}=\frac{2+\sqrt{2}}{4}$, $P_{correct}=\frac{1}{2}+\frac{1}{2\sqrt{2}}$ is maximal. Hence, the optimal error rate of distinguishing the two states is $1-P_{correct}=\frac{1}{2}-\frac{1}{2\sqrt{2}}$.

\end{document}